\documentclass[prb,floatfix,twocolumn,showpacs,amsmath,amssymb]{revtex4}
\usepackage{graphicx}
\usepackage{dcolumn}
\usepackage{bm}
\begin{document}

\title{Interaction induced Fermi-surface renormalization 
       in the $t_1{-}t_2$ Hubbard model close to the Mott-Hubbard transition}
\author{Luca F. Tocchio,$^{1}$ Federico Becca,$^{2}$
        and Claudius Gros$^{1}$ 
        }
\affiliation{
$^{1}$ Institute for Theoretical Physics, 
       Frankfurt University, 
       Max-von-Laue-Stra{\ss}e 1, D-60438 Frankfurt a.M., Germany \\
$^{2}$ CNR-IOM-Democritos National Simulation Centre 
       and International School for Advanced Studies (SISSA), 
       Via Beirut 2, I-34151, Trieste, Italy
            }

\date{\today} 

\begin{abstract}
We investigate the nature of the interaction-driven
Mott-Hubbard transition of the half-filled $t_1{-}t_2$
Hubbard model in one dimension, using a full-fledged
variational Monte Carlo approach including a 
distance-dependent Jastrow factor and backflow correlations.
We present data for the evolution of the magnetic
properties across the Mott-Hubbard transition and 
on the commensurate to incommensurate transition in the
insulating state. Analyzing renormalized excitation spectra,
we find that the Fermi surface renormalizes to perfect
nesting right at the Mott-Hubbard transition in the 
insulating state, with a first-order reorganization
when crossing into the conducting state.

\end{abstract}


\maketitle

\section{Introduction}
Low-dimensional Fermi gases and insulators with intermediate and strong 
couplings show a plethora of interesting phenomena, both in the domains of
synthesizable materials~\cite{toyota07} and of ultra-cold atom 
gases,~\cite{giorgini08} with the proximity of metallic, magnetic, 
superconducting, and insulating phases being a key target for experimental
and theoretical studies. One-dimensional correlated electron systems are 
hence good targets, to give an example, for the exploration of photo-induced 
phase transitions,~\cite{shinichiro06} having in part extremely large 
third-order non-linear optical susceptibilities, with possible applications 
to all-optical switching devices.~\cite{kishida00}

Here, we are interested in the nature of the interaction-driven Mott-Hubbard
transition which occurs in the one-dimensional $t_1{-}t_2$ Hubbard model 
at half filling. In particular, we will assess the evolution of the Fermi
surface by varying the Coulomb interaction. In the insulating state, 
the underlying Fermi-surface is given by the boundary of the occupied states 
of the renormalized dispersion relation, when the residual interactions giving
rise to the charge gap are turned off in a Gedanken 
experiment.~\cite{gros06,yoshida06,gros07,sensarma07,prelovsek}
Mathematically, the underlying Fermi surface is defined in a non Fermi-liquid 
state as the locus in $k-$space where the real part of the one-particle 
Green's function changes its sign.~\cite{gros06,dzyaloshinskii03}
By investigating magnetic and charge properties, we find that the Fermi 
surface reconstructs in a first-order manner right at the Mott transition. 
In particular, the Fermi surface is generic, namely non-nesting, in the 
metallic side, whereas it has perfect nesting properties in the insulating 
state, at the transition point.

The paper is organized as follow: in section~\ref{sec:model}, we introduce the 
Hamiltonian; in section~\ref{sec:approach}, we describe our variational wave
function; in section~\ref{sec:results}, we present our numerical results and,
finally, in section~\ref{sec:conc} we draw the conclusions.
  
\section{Model}\label{sec:model}
We consider the one-dimensional $t_1{-}t_2$ Hubbard model 
\begin{equation}\label{hubbard}
{\cal H}=-\hspace{-1ex}\sum_{i,\sigma,n=1,2} t_n c^{\dagger}_{i,\sigma} 
         c^{\phantom{\dagger}}_{i+n,\sigma} 
        + \textrm{H.c.}
        +U \sum_{i} n_{i,\uparrow} n_{i,\downarrow},
\end{equation}
where $c^{\dagger}_{i,\sigma}$ is the electron creation operator, 
$\sigma=\uparrow,\downarrow$ the electron spin, $i=1,\dots,L$ the site index, 
$n_{i,\sigma}=c^{\dagger}_{i,\sigma}c^{\phantom{\dagger}}_{i,\sigma}$
the electron density, $t_{1}$ and $t_{2}$ the nearest
and next-nearest neighbor hopping amplitudes,~\cite{signt} and $U$ the on-site
Coulomb repulsion. In this work, we focus our attention on the half-filled 
case with $L$ electrons on $L$ sites. 

\begin{figure}[t]
\includegraphics[width=\columnwidth]{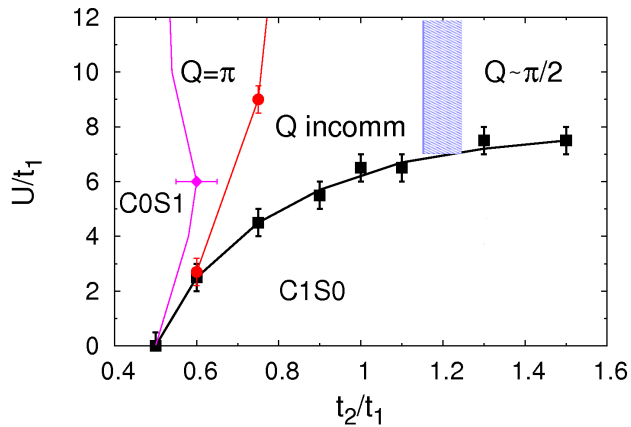}
\caption{\label{fig:phase_diagr}
(Color online) Phase diagram of the $t_1{-}t_2$ Hubbard model at half-filling
with the metallic phase with gapped spin excitations (C1S0) and the insulating
phase with gapless spin excitations (C0S1). The insulating phase with gapped
spin excitations for larger $U$ and $t_2/t_1>1/2$ has regions with commensurate
($Q=\pi$) and incommensurate ($Q$ incomm) spin-spin correlations. 
A crossover region separates the phase where the peak in $S(q)$ is 
incommensurate and the one with the peak commensurate to a doubled unit cell 
($Q\sim\pi/2$).} 
\end{figure}

The ground state of the $t_1{-}t_2$ Hubbard model at half filling is predicted
to be an insulator with gapless spin excitations (conventionally labeled as 
C0S1) for $t_2/t_1 <1/2$ and every finite $U/t_1$,~\cite{lieb} a spin-gapped 
metal (C1S0) with strong superconducting fluctuations for $t_2/t_1 >1/2$ and 
small $U/t_1$,~\cite{fabrizio} and a fully-gapped spontaneously dimerized 
insulator (C0S0) for $t_2/t_1 >1/2$ and large $U/t_1$.~\cite{arita,daul} 
Our findings, which are summarized in Fig.~\ref{fig:phase_diagr}, are in very 
good agreement with these results.
The locus of the metal-insulator transition has been investigated by several 
groups,~\cite{torio,aebischer,gros05,jap} with slightly varying outcomes.
Remarkably, a transition between incommensurate and commensurate spin 
excitations is expected to take place inside the C0S0 
phase.~\cite{daul,gros05,berthod06} Finally, we would like to mention that a
tiny C2S2 phase could be stable for $U/t_1 \to 0$, as suggested by a
weak-coupling renormalization group approach;~\cite{louis01} recent
calculations showed that this phase can be further stabilized in presence of
long-range interactions.~\cite{motrunich}

\section{Variational approach}\label{sec:approach}
In this paper, we present a variational Monte Carlo study of the Hubbard 
model for $t_2/t_1 > 1/2$, which allows us to determine accurately the locus 
of the metal-insulator transition, to study the transition between commensurate
and incommensurate spin-spin correlations in the large-$U$ (dimerized) phase 
and to investigate its underlying Fermi surface. In particular, we will show 
that the magnetic correlations are related to the single-particle spectrum in 
the optimized variational wave function. Moreover, we will propose that the 
metal-insulator transition is driven in the Mott state by a renormalization 
of the underlying Fermi surface to perfect nesting.

Both the metallic and the insulating phases can be constructed, in a 
variational approach. In a first step, one constructs uncorrelated wave 
functions given by the ground state $|\rm{BCS}\rangle$ of a superconducting 
Bardeen-Cooper-Schrieffer (BCS) Hamiltonian,~\cite{grosbcs,zhang88}
\begin{equation}
{\cal H}_{\rm{BCS}} = \sum_{q,\sigma} \epsilon_q c^{\dagger}_{q,\sigma} c^{\phantom{\dagger}}_{q,\sigma}
+ \sum_{q} \Delta_q 
c^{\dagger}_{q,\uparrow} c^{\dagger}_{-q,\downarrow} + \rm{H.c.},
\end{equation}
where both the free-band dispersion $\epsilon_q$ and the pairing amplitudes 
$\Delta_q$ are variational functions. We use the parametrization
\begin{equation}
\begin{array}{rcl}
\epsilon_q& =&  -2\tilde{t}_{1}\cos q-2\tilde{t}_{2} \cos(2q) -\mu \\
\Delta_q& =& \Delta_{1}\cos q + \Delta_{2}\cos (2q) + \Delta_{3} \cos (3q),
\end{array}
\label{def_epsilon_delta}
\end{equation}
where the effective hopping amplitudes $\tilde{t}_{1}$ and $\tilde{t}_{2}$, 
as well as the effective chemical potential $\mu$ and the local pairing 
fields $\Delta_{1}$, $\Delta_{2}$, and $\Delta_{3}$ are variational 
parameters to be optimized. The excitation spectrum for Bogoliubov 
excitations is given by 
\begin{equation}
E_q=\sqrt{\epsilon_q^2+\Delta_q^2}.
\end{equation}
The correlated state $|\Psi_{\rm{BCS}}\rangle$ is then given by
\begin{equation}\label{eq_BCS}
|\Psi_{\textrm{BCS}}\rangle = {\cal J} |\textrm{BCS}\rangle,
\end{equation}
where ${\cal J}=\exp(-1/2 \sum_{i,j} v_{i,j} n_i n_j)$ is a density-density 
Jastrow factor (including the on-site Gutzwiller term), with the $v_{i,j}$ 
being optimized independently for ever distance $|i-j|$. Notably, within this 
kind of wave function, it is possible to obtain a pure (i.e., non-magnetic) 
Mott insulator by considering a sufficiently strong Jastrow 
factor,~\cite{capello} i.e., $v_q \sim 1/q^2$ ($v_q$ being the Fourier 
transform of $v_{i,j}$) and a Luttinger-liquid wave function with arbitrary
critical exponents,~\cite{luttingerWF} whenever $v_{i,j}\sim \log|i-j|$.
In addition, a dimerized phase can be obtained just by considering a gapped
BCS spectrum $E_q$ together with $v_q \sim 1/q^2$ (that is the case
whenever $t_2/t_1> 1/2$ and $U/t_1$ is large enough).~\cite{capello}
Remarkably, in this case, finite dimer-dimer correlations are found at large
distance, even though the wave function does not break the translational
symmetry. Here, we do not report results on dimer-dimer correlations, that
are found in the C0S0 phase (see Ref.~\onlinecite{capello}), but we concentrate
on spin and charge properties, with a particular emphasis on the evolution
of the Fermi surface by changing $t_2/t_1$ and $U/t_1$.

As we demonstrated recently,~\cite{tocchio} the projected BCS state 
$|\Psi_{\rm{BCS}}\rangle$ can be improved further by considering backflow 
correlations, which modify the single-particle orbitals, in the 
same spirit as proposed by Feynman and Cohen.~\cite{feynman}
In this way, already the determinant part of the wave function includes now 
correlation effects. All results presented here are obtained by fully 
incorporating the backflow corrections and optimizing individually every 
variational parameter in $\epsilon_q$ and $\Delta_q$ of 
Eq.~(\ref{def_epsilon_delta}), in the Jastrow factor $\cal J$ of 
Eq.~(\ref{eq_BCS}), as well as backflow corrections.
 
\begin{figure}[t]
\includegraphics[width=\columnwidth]{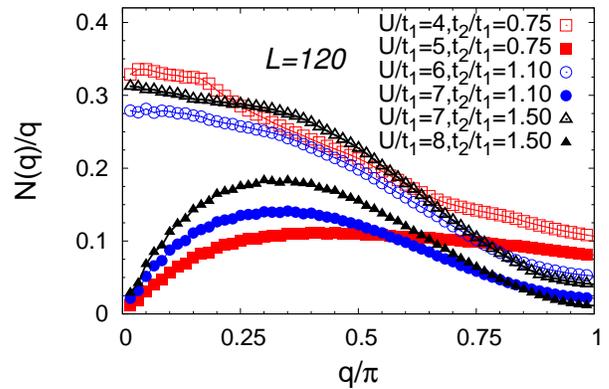}
\caption{\label{fig:n_q}
(Color online) For a chain with $L=120$ sites, the density-density correlations
$N(q)$, divided by the momentum $q$, across the metal-insulator transition for 
$t_2/t_1=0.75,1.1,1.5$. The metallic (insulating) state is characterized by 
a finite (vanishing) value of $N(q)/q$, in the limit $q \to 0$.}
\end{figure}

\section{Results}\label{sec:results}
\subsection{Mott-Hubbard transition}
The ground-state properties can be easily assessed by computing density and 
magnetic structure factors:
\begin{eqnarray}
N(q) &=& \frac{1}{L} \sum_{k,l} e^{i q (k-l)} \langle n_k n_l \rangle, \\
S(q) &=& \frac{1}{L} \sum_{k,l} e^{i q (k-l)} \langle S_k^z S_l^z \rangle,
\end{eqnarray}
where $n_k$ and $S_k^z$ are the total density and the $z$-component of the spin 
operator on site $k$, respectively.

The static density-density correlations behave qualitatively different in a 
metallic and a Mott-insulating state for small momenta $q$, with the metallic 
state being characterized by a linear dependence of $N(q) \sim q$, while in 
the insulating phase $N(q) \sim q^{2}$.~\cite{capello} In Fig.~\ref{fig:n_q},
we present the behavior of $N(q)/q$ across the transition for three values of 
the ratio $t_2/t_1$. The locus of the Mott-Hubbard transition can be determined
easily, allowing us to draw the phase diagram in Fig.~\ref{fig:phase_diagr}. 
Our determination of the line separating the metallic and the insulating phase
is in good agreement with Refs.~\onlinecite{gros05,torio}.

The metallic region in the phase diagram can be described as a Luther-Emery 
liquid, with a finite gap in the spin excitation spectrum and gapless charge 
excitations.~\cite{solyom} The charge stiffness $K_{\rho}$ can be extracted,
for example, from the long-distance behavior of the density-density
correlations. In any conducting phase, we expect that $K_{\rho}$ is also
related to the slope of $N(q)$ at small $q$, i.e., $N(q) \sim K_{\rho}|q|/\pi$. 
In fact, the latter equation, which is definitely valid in Luttinger liquids,
should hold whenever the charge degrees of freedom are gapless.~\cite{solyom}
This procedure to obtain $K_{\rho}$ works very well for the doped single-band
Hubbard model, namely for the model of Eq.~(\ref{hubbard}) with 
$t_2=0$,~\cite{capello2} in comparison with the exact results, as obtained 
by Bethe ansatz.~\cite{schulz} Unfortunately, in the metallic phase of the
$t_1{-}t_2$ model, the existence of very small gaps prevents us to obtain 
accurate results for $K_{\rho}$.

\begin{figure}[t]
\includegraphics[width=\columnwidth]{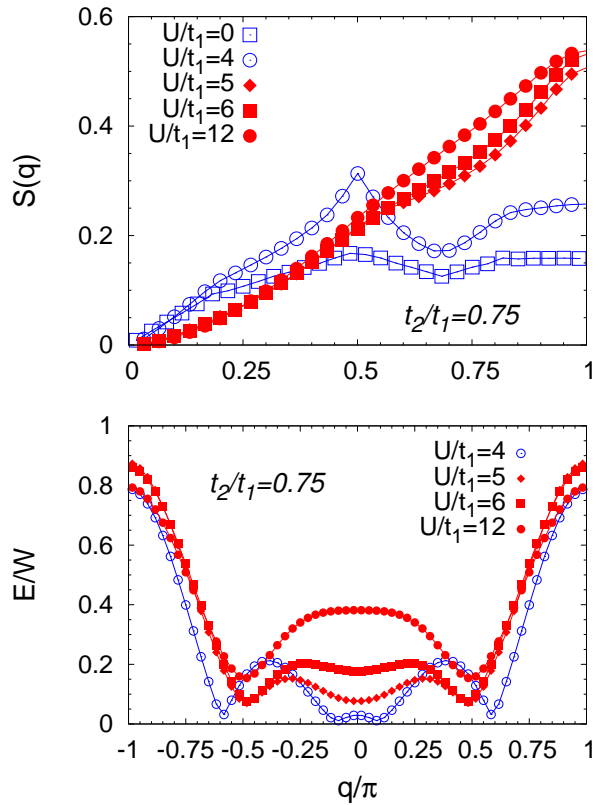}
\caption{\label{fig:spin075}
(Color online) Upper panel: Spin-spin correlations $S(q)$ at $t_2/t_1=0.75$, 
for a $L=120$ chain. Data are shown for $U/t_1=0,4$ (metal) and for 
$U/t_1=5,6,12$ (insulator). Lower panel: Single-particle spectrum $E_q/W$ at 
$t_2/t_1=0.75$ for the same values of the electron-electron repulsion $U$ and 
the same chain length.} 
\end{figure}

\subsection{Magnetic properties}
The presence of {\it short-range} magnetic order is signaled by the appearance
of a peak in $S(q)$, for a certain momentum $Q$. In the following, we will 
compare the magnetic properties with the renormalized single-particle spectrum
$E_q$ of the optimized variational wave function. The energy scale for $E_q$ 
will be taken as the bandwidth $W$ of the original free dispersion 
$\epsilon_q^0=-2t_1\cos(q)-2t_2\cos(2q)$. Note that, in the non-interacting 
case, there is only a single, perfectly nested Fermi surface for 
$t_2/t_1< 0.5$, with two Fermi points separated by $\pi$. 
Instead, for $t_2/t_1>0.5$ there are two Fermi seas and four Fermi points.

\begin{figure}[t]
\includegraphics[width=\columnwidth]{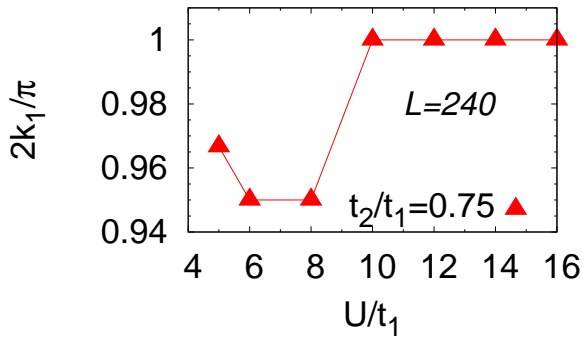}
\caption{\label{fig:fermi_vect}
(Color online) Evolution of $2k_1$ (distance between the minima of the 
single-particle spectrum $E_q$) in the insulating phase for $t_2/t_1=0.75$. 
Note the transition from an incommensurate to a commensurate value for 
$U/t_1 \simeq 9$.} 
\end{figure}

In the metallic phase, the spin properties are only slightly modified by the 
presence of a small but finite interaction $U$, with respect to the $U=0$ 
behavior; for $t_2/t_1>0.5$ the single-particle spectrum $E_q$ exhibits four 
minima, at $\pm k_{1}$ and $\pm k_{2}$, and the peak of $S(q)$ is located at 
$Q^{\rm{met}}=k_{2}-k_{1}=\pi/2$. The condition $Q^{\rm{met}}=\pi/2$ is 
determined by the Luttinger sum-rule for the metal, which states that the 
total volume of the Fermi sea equals the number of electrons.
In the insulating phase, the situation changes qualitatively and the magnetic
properties of the system become strongly affected by the value of $t_2/t_1$.

In Fig.~\ref{fig:spin075}, we show the behavior of $S(q)$ across the 
metal-insulator transition for $t_2/t_1=0.75$, in comparison with 
the variationally determined renormalized single-particle spectrum $E_q$. 
It can be observed that, when entering the insulating phase, the 
single-particle spectrum becomes strongly gapped and the two central minima 
collapse into a unique relative minimum at $q=0$, that subsequently disappears,
as $U/t_1$ increases. At the same time, the peak in $S(q)$ shifts 
from $Q^{\rm{met}}=\pi/2$ to $Q^{\rm{ins}}=\pi$. 
Remarkably, just above the Mott transition, namely for $5 < U/t_1 < 10$, the 
quantity $2k_{1}$ (i.e., the distance between the two absolute minima of $E_q$)
is slightly different from $\pi$ and becomes commensurate only after a second 
transition (e.g., $U/t_1 \simeq 9$), inside the insulating phase,~\cite{gros05}
see Fig.~\ref{fig:fermi_vect}. However, the degree of incommensurability is 
very small and does not show up in corresponding shift of the peak in 
$S(q)$ from $Q^{\rm{ins}}=\pi$. Indeed, the magnetic correlations are 
short-ranged and the peak in $S(q)$ is consequently broad. A shift of the 
momentum away from $\pi$ will therefore result in a shift of the maximum in 
$S(q)$ only for a substantial degree of incommensurability.

\begin{figure}[t]
\includegraphics[width=\columnwidth]{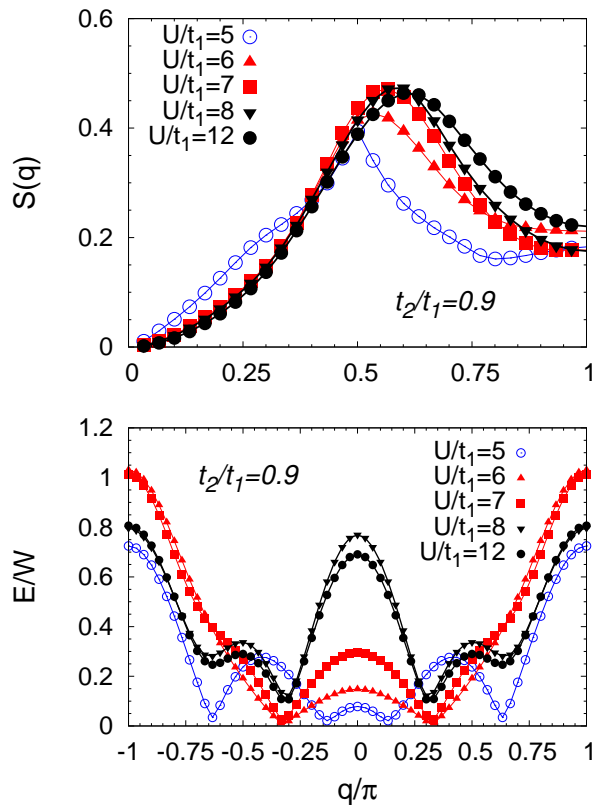}
\caption{\label{fig:spin09}
(Color online) Upper panel: Spin-spin correlations $S(q)$ at $t_2/t_1=0.9$, 
for a $L=120$ chain. Data are shown for $U/t_1=5$ (metal)
and $U/t_1=6,7,8,12$ (insulator). Lower panel: 
Single-particle spectrum $E_q/W$ at $t_2/t_1=0.9$ for the 
same values of the electron-electron repulsion $U$ and the same chain length.} 
\end{figure}

In Fig.~\ref{fig:spin09}, we plot the spin-spin correlations $S(q)$ 
and the single-particle spectrum $E_q$ for the ratio $t_2/t_1=0.9$.
In the metallic phase, the spin-spin correlations are always peaked at 
$Q^{\rm{met}}=\pi/2$, while in the insulating phase the peak slowly shifts 
to $Q\simeq 0.6\pi$. For $U=6$ and $7$ the single-particle 
spectrum $E_q$ is qualitatively different from the one for larger $U$'s. 
As shown later, this is related to the different behavior of the variational 
hopping ratio $\tilde{t}_2/\tilde{t}_1$ close to the metal-insulator 
transition with respect to the strong-coupling regime.   
For larger values of the ratio $U/t_1$, $E_q$ shows four local minima, 
with the peak in $S(q)$ located at $Q^{\rm{ins}}=2k_{1}$, where $2k_{1}$ is 
the distance between the two absolute minima.  

\begin{figure}[t]
\includegraphics[width=\columnwidth]{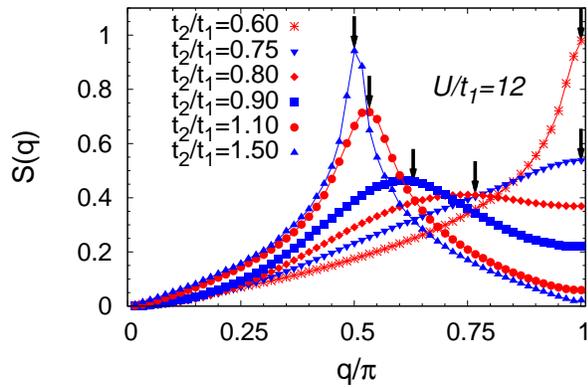}
\caption{\label{fig:spinU12}
(Color online) The spin-spin correlations $S(q)$ at $U/t_1=12$ and for $L=120$ 
sites, for different values of the hopping ratio $t_2/t_1$. Arrows indicate 
the quantity $2k_{1}$, obtained from the single-particle spectrum $E_q$.}
\end{figure}

Finally, in Fig.~\ref{fig:spinU12}, we summarize the spin-spin correlations
for different $t_2/t_1$, at a given value of $U/t_1$, chosen to be far enough 
from the metal-insulator transition in order to describe the large-$U$ 
behavior of $S(q)$. The peak in the spin-spin correlations exhibits the 
commensurate-incommensurate transition moving far from $Q=\pi$,
as the ratio $t_2/t_1$ is increased. When $t_2/t_1=1.5$ the system behaves 
already like in the $t_2/t_1\to \infty$ limit, with the peak commensurate to 
a lattice with a doubled unit cell ($Q=\pi/2$). These results are in agreement
with previous studies for the Heisenberg~\cite{DMRG} and the Hubbard 
model.~\cite{daul}

\subsection {Fermi surface renormalization}
Finally, we present our central result, namely the fact that the 
metal-insulator transition is driven, in the Mott-insulating state, by a 
renormalization of the underlying Fermi surface to perfect nesting. 
With underlying Fermi surface, we mean the locus of the highest occupied 
momenta in the non-interacting spectrum 
$\epsilon_q=-2\tilde{t}_1\cos(q)-2\tilde{t}_2\cos(2q)$, 
obtained from the optimized variational hopping parameters. 
We would like to stress that the concept of an underlying Fermi surface is 
of central importance for the angular resolved photoemission spectroscopy 
(ARPES) studies of strongly correlated systems, like the high-temperature 
superconductors.~\cite{gros06,yoshida06,gros07,sensarma07}
Note, that $E_q=\sqrt{\epsilon_q^2+\Delta_q^2}$ corresponds within 
renormalized mean-field theory~\cite{zhang88} to the excitation spectrum of 
projected Bogoliubov quasiparticles and $\epsilon_q$ hence to the dispersion 
of the renormalized quasiparticles. Moreover, recent calculations on the
$t{-}J$ and the periodic Anderson models highlighted the possibility to assess
the Fermi surface from the parametrization of a variational wave 
function.~\cite{himeda,watanabe} Here, the renormalization of the hopping
parameters made it possible to show non-trivial deformations of the 
non-interacting Fermi surface, due to the Gutzwiller projection.

\begin{figure}[t]
\includegraphics[width=\columnwidth]{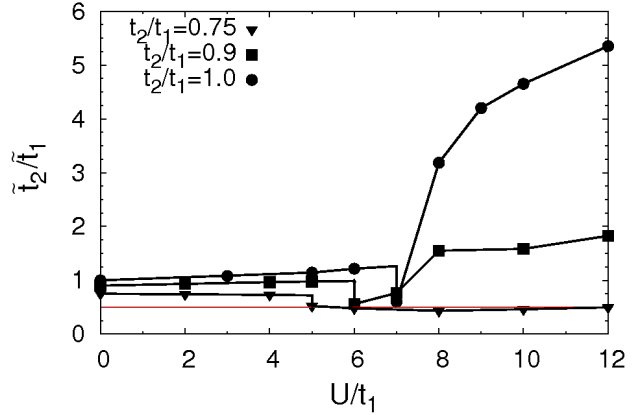}
\caption{\label{fig:ren-t}
(Color online) The variationally optimized hopping ratio 
$\tilde{t}_2/\tilde{t}_1$ in $|\textrm{BCS}\rangle$, see Eq.~(\ref{eq_BCS}), 
as a function of $U/t_1$. The metal-insulator transition takes place for 
$U/t_1 = 4.5 \pm 0.5$, $5.5 \pm 0.5$ and $6.5 \pm 0.5$ for $t_2/t_1=0.75$ 
(triangles), $0.9$ (squares) and $1.0$ (circles), respectively. 
The variational state $|\textrm{BCS}\rangle$ contains only a single Fermi 
sea for $\tilde{t}_2/\tilde{t}_1\le 0.5$ (horizontal line).}
\end{figure}

We show in Fig.~\ref{fig:ren-t} that the ratio $\tilde{t}_2/\tilde{t}_1$ in 
the metallic phase is almost equal to the bare value $t_2/t_1$, regardless 
of the degree of interaction. This weak renormalization of the band-structure 
in the metallic state is in agreement with a renormalization-group 
study,~\cite{louis01} which predicts that the renormalization of the 
Fermi surface is proportional to $U^2$. Then, after the metal-insulator 
transition, the ratio jumps to a smaller value, very close to $1/2$. 
According to our data, we propose that the optimized variational ratio of 
$\tilde{t}_2/\tilde{t}_1$ is renormalized to $1/2$, {\it exactly} at the 
metal-insulator transition. This discontinuous behavior of the renormalized 
band structure is also evident in Fig.~\ref{fig:spin09}, e.g., for 
$t_2/t_1=0.9$, with the number of minima of the single-particle spectrum $E_q$ 
jumping from four to two when entering the Mott-insulating state.
We note that an analogous tendency towards a Fermi surface symmetrization 
in the insulating state has been observed in a study of a two-dimensional 
frustrated lattice.~\cite{liu05,tocchiotriang}

A renormalization of the variational hopping ratio to 
$\tilde{t}_2/\tilde{t}_1=1/2$ implies that the Fermi surface is nested, 
with two Fermi points separated by a vector $\pi$. This perfect nesting 
condition drives the system to be an insulator, generating a charge gap as 
soon as electron-electron interaction is switched on. Remarkably, while the 
metal-insulator transition is driven by the renormalized dispersion 
$\epsilon_q$, the pairing terms $\Delta_q$ are crucial in determining 
the spin properties of the model, via the renormalized excitation spectra 
$E_q=\sqrt{\epsilon_q^2+\Delta_q^2}$. Indeed, as shown for example in 
Fig.~\ref{fig:spin09}, the minima of the single-particle spectrum at 
$t_2/t_1=0.9$ are connected by an incommensurate vector, leading to an 
incommensurate peak in $S(q)$.

\section{Conclusions}\label{sec:conc}
We have presented an extensive study of the phase diagram of the 
one-dimensional $t_1{-}t_2$ Hubbard model at half filling, with emphasis on 
the evolution of the magnetic properties and of the underlying Fermi surface 
across the interaction-driven Mott-Hubbard transition. 
We have shown that the magnetic correlations are related to the single-particle
spectrum in the optimized variational wave function and we have described how
they are affected by the metal-insulator transition. In the insulating phase, 
the peak in the spin-spin correlations exhibits the commensurate-incommensurate
transition moving far from $Q=\pi$, as the ratio $t_2/t_1$ is increased, 
and then becomes commensurate to a doubled unit cell when 
$t_2/t_1 \gtrsim 1.3$. 

Our main findings culminate in the hypothesis that the underlying Fermi surface
renormalizes to perfect nesting right at the transition in the insulating 
phase, with a first-order reorganization when crossing the transition into 
the metallic state. Similar renormalizations of the Fermi surface have been 
observed in two-dimensional models.~\cite{gros06,liu05,tocchiotriang}
Therefore, we believe that our results are important for an improved 
understanding of Mott-Hubbard transitions quite in general, 
transcending the specific one-dimensional physics.
 
L.F.T. and C.G. acknowledge the support of the German Science Foundation 
through the Transregio 49.



\begin{thebibliography}{99}
\bibitem{signt} At half-filling the relative sign between 
$t_1$ and $t_2$ is irrelevant.
\end{thebibliography}
\end{document}